\documentclass[12pt]{iopart}

\usepackage{graphicx}  
\begin{document}

\title{Voids and overdensities of coupled Dark Energy}

\author{Roberto Mainini}

\address{Institute of Theoretical Astrophysics, University of Oslo, Box 1029, 
0315 Oslo, Norway}

\begin{abstract}
We investigate the clustering properties of dynamical Dark Energy 
even in association of a possible coupling between Dark Energy and Dark Matter.
We find that within matter inhomogeneities, Dark Energy migth form voids as well
as overdensity depending on how its background energy density evolves.
Consequently and contrarily to what expected,  Dark Energy fluctuations are found to be slightly suppressed if a coupling with Dark Matter is permitted.
When considering density contrasts and scales typical of superclusters, voids
and supervoids, perturbations amplitudes 
range from $|\delta_\phi|\sim {\cal O} (10^{-6})$ to 
$|\delta_\phi|\sim {\cal O} (10^{-4})$ indicating an almost homogeneous 
Dark Energy component.   
  
\end{abstract}

\pacs{ }
\maketitle
\section{Introduction}
\label{sec:intro}

Several observations made over the recent years, related to a large extension 
to Large Scale Structures (LSS) and 
anisotropies of the Cosmic Microwave Background (CMB) 
as well as the magnitude--redshift relation for type Ia Supernovae 
\cite{tegmark}, 
have given us a convincing picture of the energy and matter density in 
the Universe.  

Baryonic matter accounts for no more than 30$\%$ of the mass in galaxy 
clusters while the existence of a large clustered component of 
Dark Matter (DM) seems now firmly established, although its nature is 
still unknown. 
However, they 
contribute to the total energy density of the Universe with 
only a few percent and about 25$\%$ respectively.

No more than another few percent could be accounted for by massive neutrinos, 
but only in the most favorable, but unlikely case. According to \cite{oystein}
the total mass of neutrinos cannot exceed the limit of 1.43 $eV$ (see, however, \cite{nu} for a recent analysis
on neutrino mass limits in coupled dark energy models). 
A very small part $(10^{-4})$ of the total energy density is due to massless
neutrinos and CMB radiation. 

The model suggested by observations is only viable if the remaining 75$\%$
is ascribed to the so--called Dark Energy (DE) responsible for the present 
day cosmic acceleration. 

Although strongly indicated by the observations, the existence of DE is even more 
puzzling than DM. It can be identified with a cosmological constant 
or with a yet unknown dynamical component with negative pressure. 
On the other hand, its manifestation can be interpreted as a geometrical 
property of the gravity on large scales resulting from a failure of 
General Relativity (GR) on those scales (see \cite{copeland} for a review). 
   
Within the context of GR, as an alternative to the cosmological constant,
DE is usually described as a self--interacting scalar field or a cosmic fluid
with negative pressure (see \cite{peebles} and references therein). It is usually assumed that density 
perturbations of DE play a negligible role in the structure formation because
of its very small mass $\sim H$ ($H$ being the Hubble parameter). 
Accordingly, perturbations should appear 
only on very large scales ($>$100 Mpc) and are bound to be linear so that rates of 
structure formation and their growth are influenced by DE only through 
the overall cosmic expansion \cite{mainini}.   

Nevertheless, this assertion remains questionable and its validity 
has been subject to recent debate. Then, the key question is whether
DE actively participates in the clustering and virialization 
processes possibly developing non--linearity on relevant scales. 
Some attempts to solve the problem have been done in \cite{matarrese,nunes,
dutta,mota},
sometimes adopting a phenomenological approach parametrizing the 
observables associated with DE clustering or in the context of coupled DE
and other non--trivial DE models where clustering is expected to be more 
probable.
 
An intriguing result was obtained by Dutta \& Maor \cite{dutta}.
They numerically studied the clustering of DE within matter overdensities 
showing that DE tends to form underdensities or voids in response to
the gravitationally collapsing matter (similarly, DE overdensities are expected in correspondence of matter underdensities). On supercluster scales they found 
$|\delta_{\phi}|\sim {\cal O}(10^{-2})$ when $\delta_m\sim {\cal O}(1)$ which 
could be relevant to observations 
($\delta_m$ and $\delta_{\phi}$ being the density contrast of matter an DE 
respectively).   
However, their analysis was limited to linear spherical perturbations
and the simplest class of DE models in which DE is ascribed 
to a ligth scalar field $\phi$ slowly--rolling down its pontential $V(\phi)$, 
minimally coupled to matter and gravity. 

A similar problem was investigated by Mota et al \cite{mota} with a different approach   
providing an analytical approximation for $\delta_{\phi}$ valid both in linear 
and nonlinear regime ($\delta_m > {\cal O}(1)$) in uncoupled scalar field 
DE models. However,
in spite of the qualitative agreement of their results with 
those of Dutta \& Maor \cite{dutta}, they found 
$|\delta_{\phi}|\sim {\cal O}(10^{-5})$ when the same linear scales are 
considered.

In the present paper we extend the analysis of Dutta \& Maor \cite{dutta} taking into account
a possible coupling between DM and DE and considering a wider class of DE tracking potentials. 
Then, we numerically study DE clustering on those
scales for which $\delta_m \sim {\cal O}(1)$ today, i.e. supercluster, void and supervoid scales,
and show that, coupled as well as uncoupled DE develops inhomogeneities which 
amplitudes  are consistent with the findings of Mota et al \cite{mota}. Formation of DE voids (overdensities) is however related to how the background energy density of DE evolves and not only to the presence of matter overdensities (underdensities) as claimed in the previous works.
It is also shown that, as expected, the growth of matter fluctuations is suppressed 
as soon as DE starts to drive the cosmic acceleration while 
$\delta_{\phi} \rightarrow 0$ as DE, in the models here considered,  
asymptotically approaches a cosmological constant.
 
The plan of the paper is as follows: in Section \ref{model} we describe 
our model and give the linearized equations for matter, DE and 
metric perturbations which
are derived in  \ref{equations}. Numerical results concerning perturbation 
evolution are presented and discussed in Section \ref{results}.
We summarize and conclude in Section \ref{conclusion}.

\section{The model}
\label{model}

The essential feature of a scalar field $\phi$, in order to yield DE
and drive the cosmic acceleration, is 
its self--interaction through a potential $V(\phi)$.

In addition to self--interaction, a scalar field can in principle be 
coupled to any other field present in nature.
Couplings to ordinary particles are strongly constrained by observational 
limits on violations of the equivalence principle
but limits on the DM coupling are looser (constraints on coupling
for specific models were obtained in \cite{maccio,constraints} from CMB,
N-body simulations and matter power spectrum analysis).  
If present, DM coupling could have a relevant role in 
the cosmological evolution affecting not only the overall cosmic expansion
but also modifying the DM particles dynamics with  relevant 
consequences for the growth of the density perturbations in both the 
linear and nonlinear regime (e.g., on halo density profiles, 
mass function and its evolution)  \cite{maccio,amendola,dolag}.
Here we consider one of the most popular models where a coupling between DM and 
DE is present, namely coupled DE \cite{amendola} (for different DE--matter 
interactions see \cite{interaction}). 

According to general covariance, the sum of the individual stress energy tensors
$T^\mu_{(i) \nu}$ ($i=b,dm,\phi$) must be conserved so we can write:

\begin{equation}
\fl
\nabla_{\mu} T^{\mu}_{(b) \nu} = 0\\
\nabla_{\mu} T^{\mu}_{(dm) \nu} = -CT_{(dm)}\nabla_\nu\phi\\
\nabla_{\mu} T^{\mu}_{(\phi) \nu} = CT_{(dm)}\nabla_\nu\phi
\label{contin}
\end{equation}
Here $\nabla_{\mu}$ is the covariant derivative, 
$T_{(dm)}$  indicates the trace of the DM stress energy tensor
and $C$ is a constant which parametrizes the strength 
of the DM--DE interaction.

In a spacetime described by a metric $g_{\mu\nu}$  
with signature $(-+++)$, the DE stress energy tensor takes 
the form:
\begin{equation}
\label{Tde}
T^\mu_{(\phi)\nu} = \partial^\mu \phi \partial_\nu \phi - \delta^\mu_\nu \left(
{1\over 2} \partial_\sigma \phi \partial^\sigma \phi +V\right) 
\end{equation}
while baryons and DM are well described as non--relativistic pressureless 
perfect fluids with:
\begin{equation}
T^{\mu}_{(i) \nu} =  \rho_i  u^{\mu}_{(i)} u_{\nu (i)}
\end{equation}
where $\rho_i$ is the energy density of the component $i=b,dm$ and 
$u^\mu_{(i)}$ is its 4--velocity.

As we are interested in spherical perturbations around a spatially flat 
Friedmann--Robertson--Walker (FWR) background, only spherically
symmetric spacetimes are considered for which the most general line element 
in comoving coordinates is:
\begin{equation}
\label{metric}
	ds^{2}=-dt^2+{\cal U}(t,r)dr^2+{\cal V}(t,r)\left(
	d\theta^2+\sin^{2}\theta d\varphi^2 \right) ~,
\end{equation}
where ${\cal U}(t,r)$ and ${\cal V}(t,r)$ are general functions \cite{example}.

\subsection{Background and perturbation equations}
As usual, equations for background and perturbation evolution follow
from (\ref{contin}) and Einstein's equations.
In this Section we only state the full set of equations while
their derivation is detailed in \ref{equations}.
Working in the synchronous gauge, we redefine the metric functions as follows:
\begin{eqnarray}
  {\cal U}(t,r) = a(t)^2 e^{2\zeta(t,r)} \nonumber \\
  {\cal V}(t,r) = r^2 a(t)^2 e^{2\psi(t,r)} 
 \label{redef}
\end{eqnarray}
Here $a(t)$ is the scale factor of the homogeneous 
background while $\zeta$ and $\psi$ represent deviations from homogeneity.
In the following, however, metric perturbations
will be described by $\chi=\dot\zeta+2\dot \psi$, the only 
combination which is relevant to the equations of motion (see \ref{equations} 
for more details).  
We also decompose $\phi$, $\rho_{dm}$ and $\rho_b$ as the sum of an 
unperturbed part, denoted by $ \bar{}~$, and a perturbed one:
 \begin{eqnarray}
    \phi(t,r)={\bar \phi}(t)+\delta\phi(t,r) \nonumber \\
    \rho_{dm}(t,r)={\bar \rho}_{dm}(t)+\delta\rho_{dm}(t,r) \nonumber \\
    \rho_b(t,r)={\bar \rho}_b(t)+\delta\rho_b(t,r)
\label{split} 
\end{eqnarray}
and, for each component $i$, we define the density parameter 
$\Omega_i={{\bar\rho}_i/ \rho_{cr}}$, 
the density contrast 
$\delta_i={\delta\rho_i /{\bar\rho}_i}$
and 
$\theta_i=div~{\bf v}_i$ 
($\rho_{cr}$ and ${\bf v}_i$ being the 
critical energy density and the coordinate velocity respectively). We assume 
perturbations to be small so that linear approximation applies. Further,  
only radial motions are considered, i.e. $|{\bf v}|=v^r=dr/d\tau$.

Hereafter we make use of conformal time $\tau$ related to cosmic time $t$ 
via the equation $dt=ad\tau$. Derivatives with 
respect to $\tau$ are denoted with an overdot and the conformal Hubble 
function ${\cal H}=\dot a/a = a H$.

From (\ref{Tde}), it is possible to work out the
background energy density of DE, its pressure
${\bar P}_\phi$, the corresponding perturbations as well as the 
radial velocity:
\begin{eqnarray}
\nonumber
{\bar \rho}_\phi= {\dot {\bar \phi^2} \over 2 a^2} +{\bar V} ~~~~~~
\delta\rho_\phi= {\dot {\bar \phi} \dot {\delta\phi} \over a^2}
+{\bar V'}\delta\phi\\
\nonumber
{\bar P}_\phi= {\dot {\bar \phi^2} \over 2 a^2} -{\bar V} ~~~~~~
\delta P_\phi= {\dot {\bar \phi} \dot {\delta\phi} \over a^2} - 
{\bar V'}\delta\phi  \\
v^r_\phi= {\partial_r \delta\phi \over \dot {\bar\phi}}
\label{def}
\end{eqnarray}
where ${\bar V}=V({\bar \phi})$ and $'$ denotes the derivative 
with respect to $\bar\phi$.

With the above notation and definitions, the equations for background evolution 
read:
\begin{eqnarray}
\nonumber
  3{\cal H}^2= 8\pi G \left[{\bar\rho}_{dm}+ {\bar\rho}_b+ 
{\bar\rho}_\phi\right]a^2\\
\nonumber
  \ddot{\bar\phi}+2 {\cal H}\dot{\bar\phi}+a^2{\bar V'}= C{\bar\rho}_{dm} a^2\\
\nonumber
 \dot{\bar\rho}_{dm}+3{\cal H}{\bar\rho}_{dm}= 
-C{\bar\rho}_{dm}\dot{\bar \phi} \\
  \dot{\bar\rho}_{b}+3{\cal H}{\bar\rho}_{b}=0  
\label{eqback}
\end{eqnarray}
while linear perturbations evolve according to: 
\begin{eqnarray}
\nonumber
\ddot \delta\phi + 2{\cal H} \dot \delta\phi - \nabla^2 \delta\phi +
a^2 {\bar V''}\delta \phi + \chi \dot {\bar\phi}= C\delta_{dm} 
{\bar\rho}_{dm} a^2 \\
\nonumber
\dot \delta_{dm} + \theta_{dm} + \chi= -C\dot{\delta\phi} \\
\nonumber
\dot \theta_{dm} +{\cal H} \theta_{dm} = C
\left( \dot{\bar\phi}\theta_{dm}+\nabla^2\delta\phi \right)\\
\nonumber
\dot \delta_b + \theta_b + \chi=0\\
\nonumber
\dot \theta_b +{\cal H} \theta_b = 0 \\
\dot \chi + {\cal H}\chi - {3 \over 2} {\cal H} ^2 \left[ \Omega_{dm} \delta_{dm}
+\Omega_b \delta_b + \Omega_\phi\left(\delta_\phi+
3{\delta P_\phi \over {\bar\rho}_\phi}
\right) \right] =0 
\label{eqpert}
\end{eqnarray}
Note that, in the absence of coupling, $C=0$, DM particles 
and baryons follow the same dynamics and can be used to define the 
synchronous coordinates and therefore have zero peculiar velocity 
($\theta_{dm}=\theta_b=0$). $\theta_i$'s equations are not needed anymore
and the set of equations (\ref{eqback}) and (\ref{eqpert}) then reduces to 
that given in Dutta \& Maor \cite{dutta}. 

On the other hand, as widely discussed in \cite{amendola},  
coupling modifies the dynamics of DM particles and an 
equation for $\theta_{dm}$ is therefore needed: 
although $\theta_{dm}=0$ initially, it can not remain null for all times
because of the term 
$C \left(\nabla^2\delta\phi +\dot{\bar\phi}\theta_{dm}\right)$.
As a consequence baryons and DM develop a bias $b$, i.e. 
$\delta_b=b \delta_{dm}$.
Anyway, it is still possible to eliminate the variable $\theta_b$
reducing the number of equations by one.

It is also worth mentioning that, unlike the uncoupled case, in the presence of 
coupling, Universe goes through
an evolutionary phase named $\phi$--matter dominated era ($\phi$MDE) just after 
matter--radiation equivalence. In this period
the scalar field $\phi$ behaves as $\it stiff ~matter$ ($P_\phi / \rho_\phi=1$)
having a non--negligible kinetic energy which dominates over the potential one. 
After this stage, the usual matter era follows  before entering in the 
accelerated regime.

In order to perform a numerical integration, it is most convenient to work 
in the Fourier space. Setting $\theta_b=0$, equations (\ref{eqpert}) then become:
\begin{eqnarray}
\nonumber
\ddot \delta\phi_k + 2{\cal H} \dot \delta\phi_k +k^2 \delta\phi_k +
a^2 {\bar V''}\delta \phi_k + \chi_k \dot {\bar\phi}= C\delta_{dm,k} 
{\bar\rho}_{dm} a^2 \\
\nonumber
\dot \delta_{dm,k} + \theta_{dm,k} + \chi_k= -C\dot{\delta\phi_k} \\
\nonumber
\dot \theta_{dm,k} +{\cal H} \theta_{dm,k} = C
\left(\dot{\bar\phi}\theta_{dm,k}-k^2\delta\phi_k \right)\\
\dot \delta_{b,k} + \chi_k=0\\
\nonumber
\dot \chi_k + {\cal H}\chi_k - 
{3 \over 2} {\cal H} ^2 \left[ \Omega_{dm} \delta_{dm,k}
+\Omega_b \delta_{b,k} + \Omega_\phi\left(\delta_{\phi,k}+
3{\delta P_{\phi,k} \over {\bar\rho}_\phi}
\right) \right] =0 
\label{eqpertk}
\end{eqnarray}
where the index $k$ denotes the Fourier--components with wavenumber $k$.

After having numerically evolved the above equations, their solutions are then Fourier transformed back to the coordinate space.

\subsection{Potential}
The present analysis is based on the assumption that DE is a
self--interacting scalar field $\phi$. Two self--interaction
potentials are considered \cite{rp}:  
\begin{equation}
\nonumber
V(\phi) = \Lambda^{\alpha+4}/\phi^\alpha~~~~~~~~~~~~~~~~~~~~~~~~~~~~~~~~~~~~~~~~RP
\label{RP}
\end{equation}
or
\begin{equation}
\nonumber
V(\phi) = (\Lambda^{\alpha+4}/\phi^\alpha) \exp(4\pi {\phi^2 \over m_p^2})
~~~~~~~~~~~~~~~~~~SUGRA
\label{sugra}
\end{equation}
($m_p=G^{-1/2}$ is the Planck mass) admitting tracker solutions and 
initially introduced to ease the
fine tuning and coincidence problems.
In the absence of DM--DE coupling, RP yields quite a slowly varying 
$w(a)={\bar P}_\phi / {\bar\rho}_\phi$ state parameter. On the contrary, 
SUGRA yields a fastly varying $w(a)$.
Although coupling causes a $w(a)$ behavior significantly different from the 
uncoupled case, one could again consider these potentials as examples 
of rapidly or slowly varying $w(a)$. 

For any choice of the energy scale $\Lambda$ and the positive parameter $\alpha$, the above potentials
yield a fixed $\Omega_{\phi}$. Here we prefer
to use $\Lambda$ and $\Omega_{\phi}$ as free parameters; the
related $\alpha$ value is then suitably fixed. $\Lambda$ values are chosen according to the constraints
given in \cite{gervasi}.

The use of the above potentials also permit to verify the generality of the 
Dutta \& Maor \cite{dutta} results obtained for different ones, namely the mass potential 
$V=m^2\phi^2/2$ and the double exponential potential \cite{barreiro}.

In both uncoupled and coupled cases, the above potentials admit a final de Sitter attractor ($\Omega_\phi=1$, $w=-1$) toward which universe asymptotically evolves. The behavior of the scalar field in this last stage is however different in the two cases. In the SUGRA model, driven by damped oscillations, the field
approaches the minimun of its potential rather than asymptotically  approaching infinity as in the RP case.
As we will see in the next section, perturbations reflect the same behaviors.

\subsection{Initial conditions} 

As in Dutta \& Maor \cite{dutta}, perturbations in baryons and DM are initially taken to 
be gaussian:
\begin{equation}
\delta_{dm}(\tau_i,r)=A(\tau_i)e^{-{r^2\over \sigma^2}} ~~~~~~~~ \delta_b(\tau_i,r)
=b \delta_{dm}(\tau_i,r)
\end{equation}
(here $\tau_i$ is some initial time in the matter era).
The ratio between fluctuation amplitudes in baryon and DM is prescibed by  
linear theory \cite{amendola}:
\begin{equation}
b \simeq
{ 3 \Omega_{dm} \over
3 \gamma\, \Omega_{dm} + 4\beta \mu \sqrt{\Omega_k}} 
\end{equation}
where  $\beta = \sqrt {3/16 \pi} m_p C$ is the adimensional coupling parameter, $\mu = (\dot\delta_{dm}/\delta_{dm})/(\dot a/a)$, $\gamma = 1+4\beta^2/3$
and $\Omega_k = \dot \phi^2/2a^2 \rho_{cr}$. 

The initial perturbation amplitude $A(\tau_i)$ is chosen such that the mean
value of the total matter density contrast 
$\delta = {(\delta\rho_b+\delta\rho_{dm}) / (\rho_b+\rho_{dm}}) $
within the comoving radius $R=\sigma$ at the present time $\tau_0$ is: 
\begin{equation}
|\bar \delta (r<R,\tau_0)| = |{\int^R_0 \delta (r,\tau_0)r^2 dr 
\over \int^R_0 r^2 dr}| \sim 1
\end{equation}
when dealing with supercluster and voids scales ($\sim 10-25h^{-1}$ Mpc). 
Slightly smaller $|{\bar\delta}_m|$ will be considered for supervoid scales
(see next section). 

We assume the shape of matter perturbations to be only slightly affected 
during their evolution so that $\delta_{dm}$  and $\delta_b$ are still  
well approximated by a gaussian at $\tau_0$.
We will see in the next section that 
this assumption is confirmed by numerical results.
Thus, setting $\delta= N\delta_{dm} = NAe^{-{r^2\over \sigma^2}} $we have:
\begin{equation}
\bar \delta (r<R,\tau) ={3\over 2} NA
\left[{\sqrt{\pi}\over 2} {\sigma \over
R} erf \left({R\over \sigma}\right) - e^{-{R^2\over \sigma^2}}\right]
{\sigma^2 \over R^2}
\label{delmedio}
\end{equation}
where $N ={(b~\Omega_b+\Omega_{dm}) / (\Omega_b+\Omega_{dm}})$.
For $R=\sigma$ and $\tau=\tau_0$ it follows $A\sim {1.76 / N}$ when 
${\bar \delta}\sim 1$.

\begin{figure}[h!]
\centering
\includegraphics[height=10.cm,angle=0]{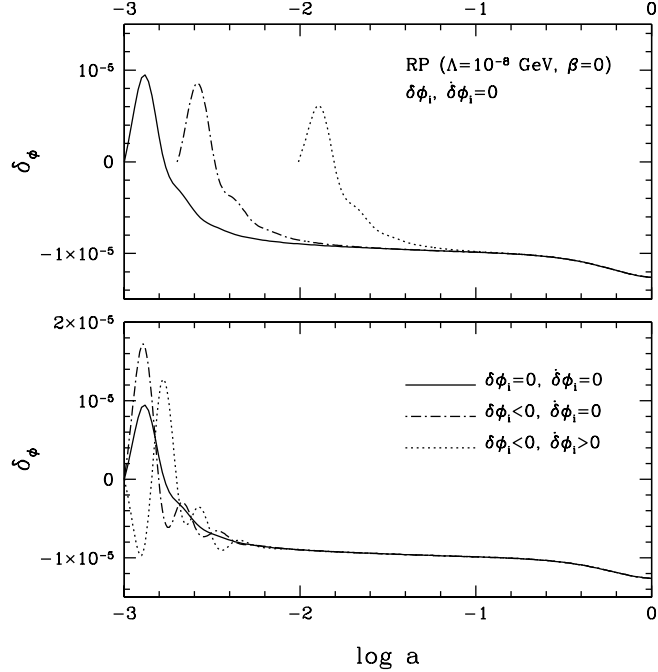}
\caption{Evolution of the DE density contrast, $\delta_\phi$, at the center 
of a spherical matter overdensity for different initial conditions.
In the top panel we set $\delta\phi$, $\dot{\delta\phi}=0$ at different 
initial redshifts ($z_i=100,500,1000$) while in the bottom panel we plot
the behavior of $\delta_\phi$ when the initial values of  
$\delta\phi$ and $\dot{\delta\phi}$ differ from zero. 
Independently from the initial conditions chosen, after an 
initial transient period, $\delta_\phi$  
always settles on the same (tracker) solution.}
\label{fig0}
\end{figure}

Dutta \& Maor \cite{dutta} set the initial redshift to $z_i=35$ assuming $\delta\phi, \dot {\delta\phi} =0$.
However, DE fluctuations would have evolved since the earlier stages of the Universe and, 
although small, they are expected to differ from zero at $z_i$ being then in tracking regime. 
With their choice for the initial conditions, 
Dutta \& Maor \cite{dutta} find that, initially, DE has a weak tendency to collapse. 
However, this effect has to be understood  as due to the fact that such initial values  do not lie 
on  the tracker solution of the perturbed scalar field equation. This is clearly illustrated
in Figure \ref{fig0} where the evolution of the density contrast at the center of a matter overdensity is plotted for different choices of the initial conditions.
In the top panel we compare the behaviors of $\delta_\phi$ when choosing 
$\delta\phi, \dot {\delta\phi} =0$  at different initial redshifts, i.e. 
$z_i=100,500,1000$. Notice the initial tendency of DE to cluster. 
However, for different choices of $\delta\phi$ and $\dot {\delta\phi}$ the initial
behavior of $\delta_\phi$ will be completely different. This is shown
in the bottom panel, where we permit initial values to differ from zero. 
An inspection of the plots, shows that independently from the initial 
conditions 
chosen, after an initial transient period, $\delta_\phi$  
always settles on the same (tracker) solution. Plots are given for a specific
RP model but the
same conclusions are reached when considering different cases.

Based on the above considerations, we choose initial conditions such that $\delta\phi$ arranges itself on the tracker solution already at $z=100$ (e.g. $\delta\phi$, $\dot{\delta\phi}=0$, $z_i=1000$). 

Finally, we set $\chi(\tau_i,r)=0$ assuring that matter perturbations initially expand with the Hubble flow.

\section{Results}
\label{results}

\begin{figure}[h!]
\centering
\includegraphics[height=10.cm,angle=0]{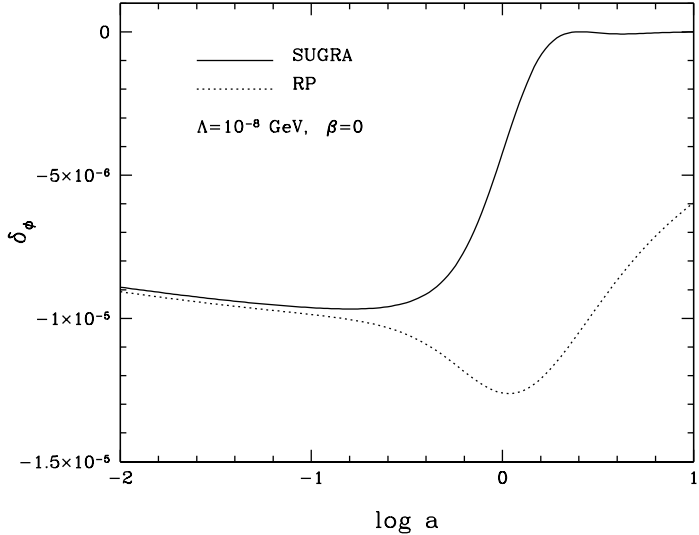}
\caption{Evolution of the DE density contrast, $\delta_\phi$, at the center 
of a spherical matter overdensity in RP (dotted line) and SUGRA (solid line) 
uncoupled models. The comoving scale of the perturbation is 
$\sigma=20h^{-1}$ Mpc and the mean matter density contrast within the 
radius $R=\sigma$ is ${\bar\delta}=1$  at the present time.}
\label{fig1}
\end{figure}

In this section we present the results of numerical runs. In order 
to better understand the behavior of the perturbations around the present
time, we stopped our runs at some time in the future. Results are then 
shown in the redshift range $100<z<-0.99$ and, for
$\sigma=20h^{-1}$ Mpc if not otherwise specified. All models considered
are spatially flat, have adimensional Hubble parameter $h=0.70$,
$\Omega_b= 0.046$ and $\Omega_{dm}= 0.234$.  

Let us start first with the uncoupled case. 
Figure \ref{fig1} compares the time evolution of the DE density contrast, 
$\delta_\phi$, 
at the center of matter overdensity ($r=0$), in  RP and SUGRA (uncoupled) 
cases. Both models present the same qualitative behavior until DE starts 
to dominate the cosmic expansion and $\phi$ approaches the Plack mass $m_p$.
Differences in the late time behaviors are then to ascribe to the exponential 
term in (\ref{sugra}) which dictates the late evolution of
$\phi$ in SUGRA model as explained in the previous section. 

Just as Dutta \& Maor \cite{dutta} and  Mota et al \cite{mota} we find that DE tends to form voids in 
correspondence of matter overdentisities obtaining 
$|\delta_\phi|\sim {\cal O}(10^{-5}-10^{-6})$ in late matter era.
While this is consistent with the findings of  Mota et al \cite{mota}, it does not 
agree with those reported by Dutta \& Maor \cite{dutta}, i.e. $|\delta_\phi|\sim {\cal O}(10^{-2})$.
\begin{figure}[h!]
\centering
\includegraphics[height=10.cm,angle=0]{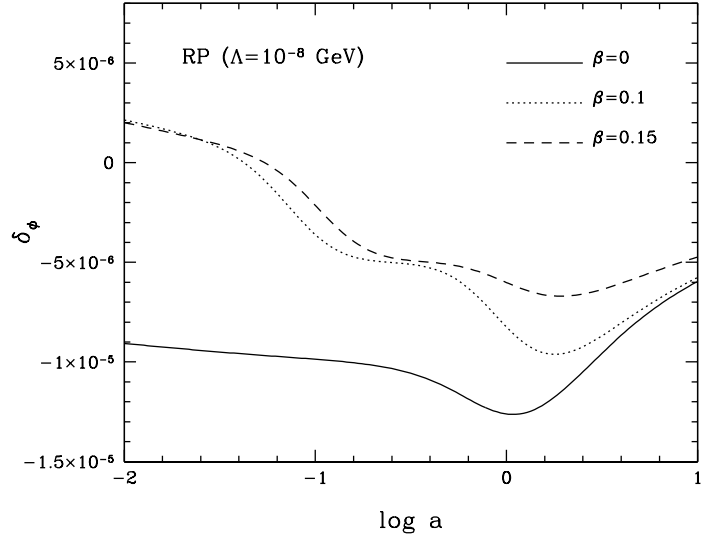}
\vskip -1.1truecm
\caption{Evolution of the DE density contrast, $\delta_\phi$, at the center 
of a spherical matter overdensity  for different values of the 
coupling parameter $\beta$ in RP models. 
The comoving scale of the perturbation is 
$\sigma=20h^{-1}$ Mpc and the mean matter density contrast within the 
radius $R=\sigma$ is ${\bar\delta}=1$  at the present time.
}
\label{fig2}
\end{figure}
\begin{figure}[h!]
\centering
\includegraphics[height=10.cm,angle=0]{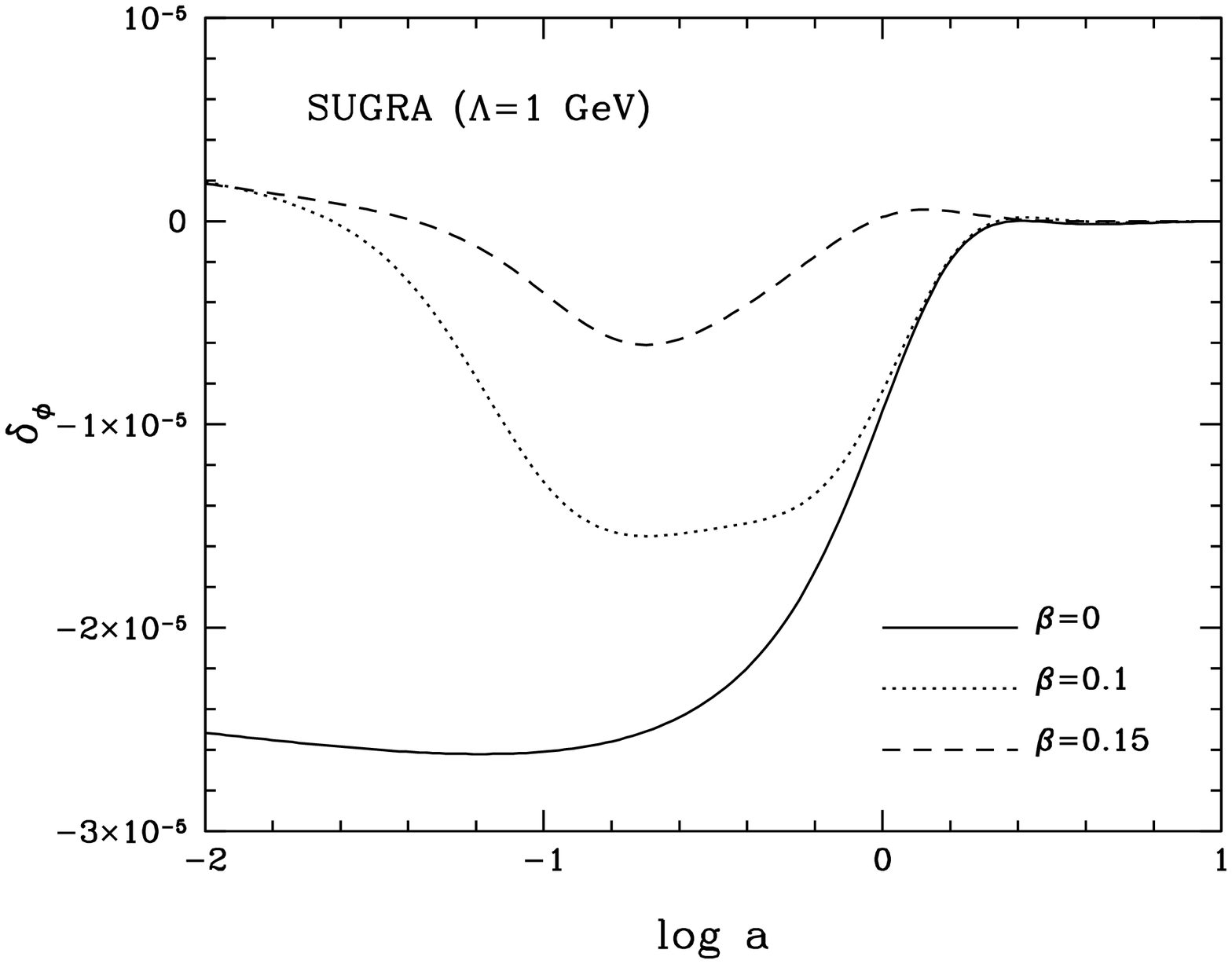}
\caption{Same as Figure \ref{fig2} but using the SUGRA potential.}
\label{fig3}
\end{figure}
\begin{figure}[h!]
\centering
\includegraphics[height=10.cm,angle=0]{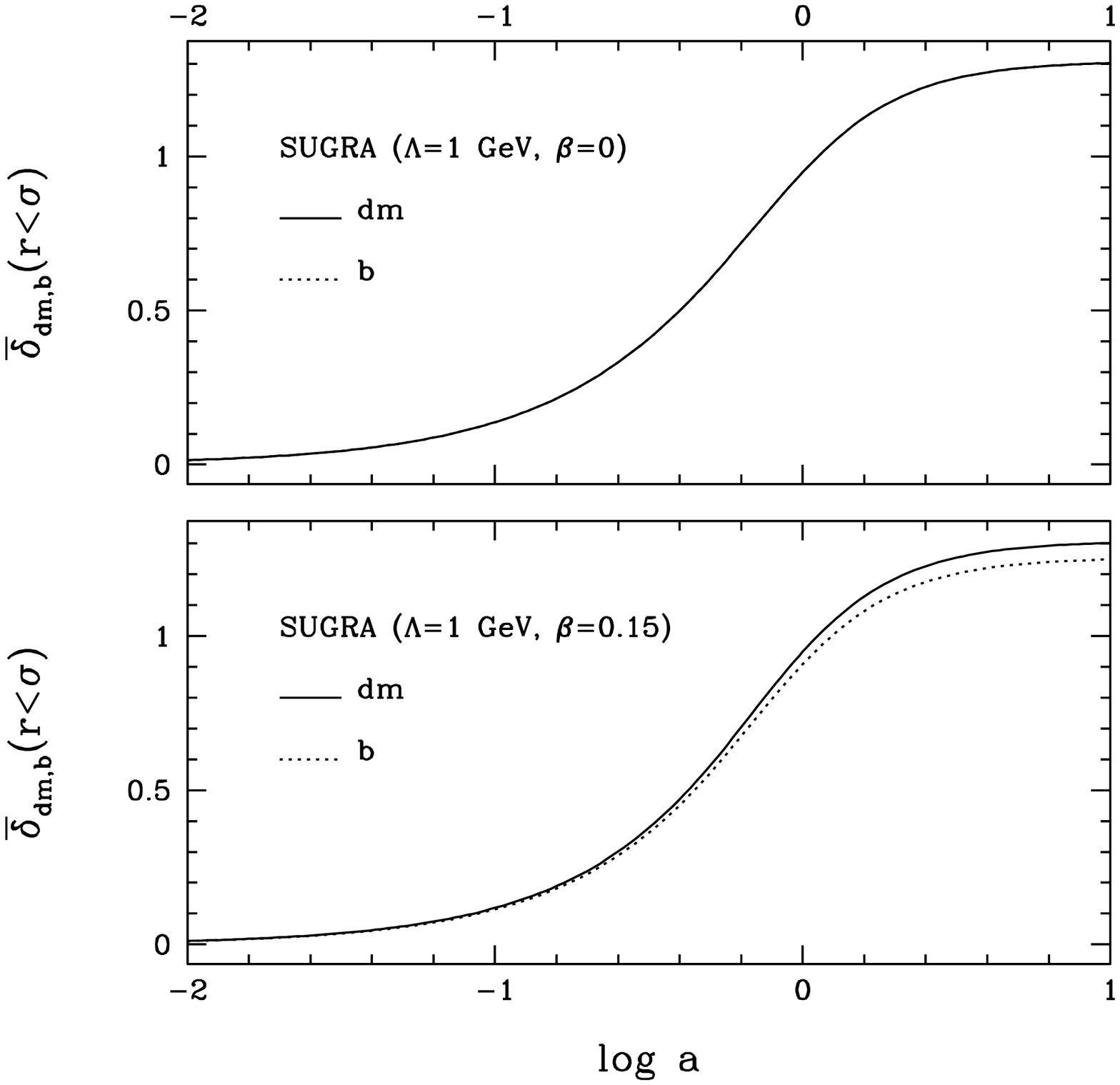}
\caption{{\it Top panel}: evolution of the mean density contrast, 
${\bar \delta_{dm,b}}$ of DM (solid line) and baryons (dotted line) in the
uncoupled SUGRA case. {\it Bottom panel}: same as top panel but in the coupled 
case. Notice how the coupling introduces a bias between DM and baryons causing 
fluctuations to grow at different rates.
Plots are given for the comoving scale $\sigma=20h^{-1}$ Mpc.
The mean matter density contrast within the radius $R=\sigma$ is 
${\bar\delta}=1$ at the present time.} 
\label{fig4}
\end{figure}
Following the same arguments by Dutta \& Maor \cite{dutta}, voids formation is understood  as a
'drag effect' due to the slower expansion of the regions with matter
overdensities. Matter collapse lowers the local value of 
$H$, reducing the Hubble damping to the scalar field. 
Therefore, in those regions, $\phi$ rolls down its potential
sligthly faster increasing its background value ${\bar \phi}$ and
$\dot {\bar\phi}$ by the quantities $\delta\phi$ and $\dot \delta\phi$ respectively. 
Noticing that ${\bar \phi}$, $\dot {\bar\phi}$, $\delta\phi$, 
$\dot \delta\phi > 0$ while ${\bar V'<0}$, a local void in DE, i.e. 
$\delta \rho_{\phi}=\delta\rho_V+\delta\rho_k<0$, will form when the local 
variation in the potential energy density, $\delta\rho_V={\bar V'}\delta\phi$, 
dominates over that in the kinetic energy density 
$\delta\rho_k=\dot {\bar\phi}\dot \delta\phi /a^2$.

This is exactly what happens in the uncoupled models here considered
during the tracking regime in matter era. In fact,
in order to have $|\delta\rho_V|>|\delta\rho_k|$ it must be: 
\begin{equation}
| {\delta\rho_V \over \delta\rho_k}|= {\alpha \over 2} {{\bar\rho}_V
\over {\bar \rho}_k}= {\alpha \over 2}{1-w \over 1+w} > 1
\label{condition}
\end{equation}
where ${\bar \rho}_k=\dot{\bar \phi^2}/2a^2$ and ${\bar\rho}_V= {\bar V}$
are the kinetic and potential background energy densities of DE and we made use 
of the tracker solution ${\bar\phi}\propto \delta\phi \propto
\tau^{6/\alpha+2}$. Noticing that in this regime we also have 
$w=-2/(\alpha+2)$, 
relation (\ref{condition}) is always satisfied requiring $\alpha>0$.

The above arguments, however, show that in general a local excess
of matter is necessary but not sufficient to assure the 
formation of a corresponding DE underdensity, the mechanism being related
to the behavior of the background energy density of DE, e.g. in our specific 
case, through the relation:
\begin{equation}
{\bar\rho}_V > {2\over \alpha} {\bar\rho}_k
\label{relation}
\end{equation}

Therefore, some differences in the evolution of $\delta_\phi$ arise when DM--DE 
coupling is considered, mainly due to the presence of the $\phi$MDE. 
As long as the $\phi$MDE holds, the kinetic energy of DE 
dominates over its potential energy invalidating (\ref{relation})
and yielding a positive $\delta_\phi$.
This is shown in Figures \ref{fig2} and \ref{fig3} which plot
the time evolution of $\delta_\phi$ for different values of the coupling 
parameter $\beta$ in RP and SUGRA cases. 

\begin{figure}[h!]
\centering
\includegraphics[height=7.cm,angle=0]{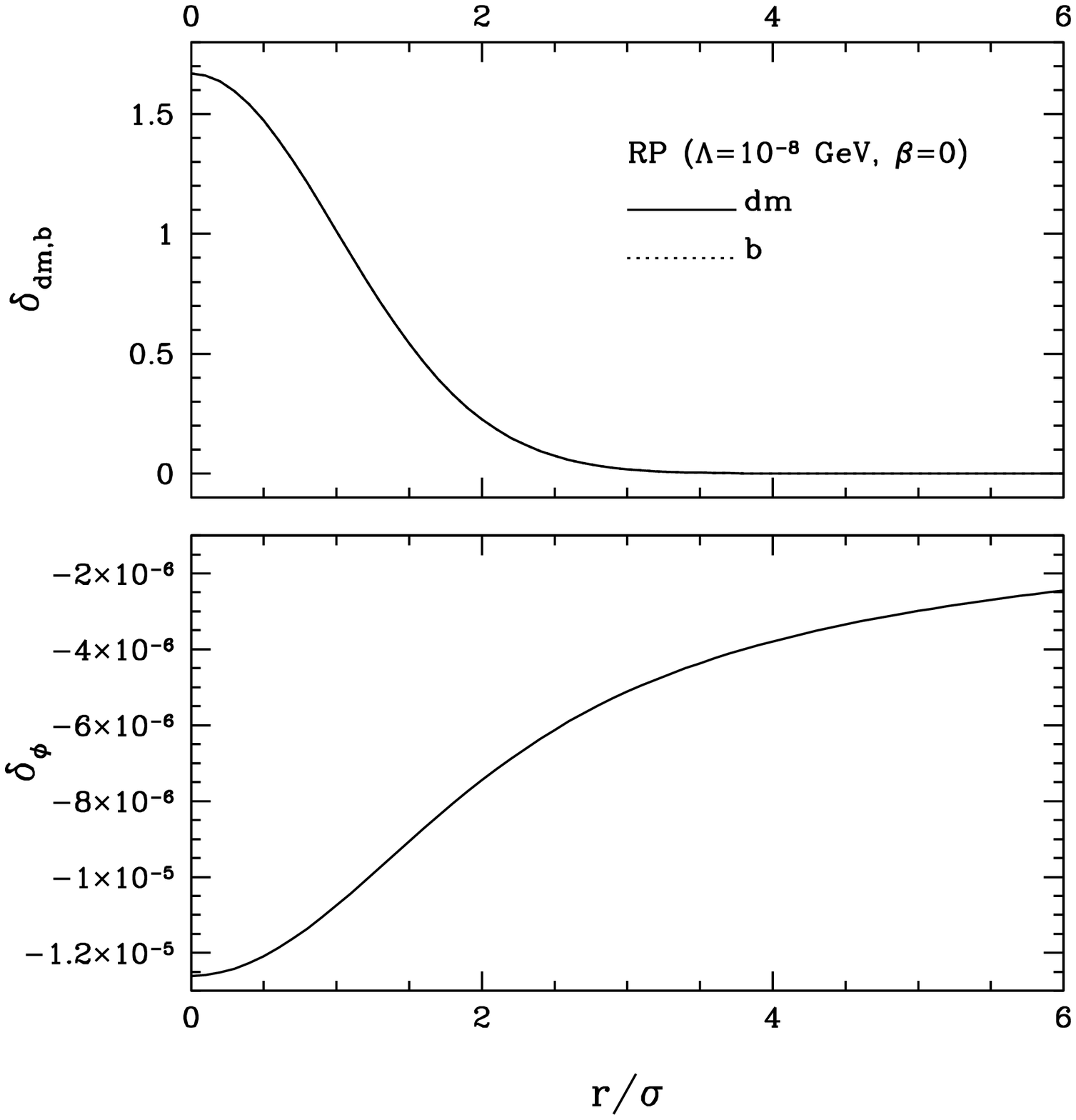}
\includegraphics[height=7.cm,angle=0]{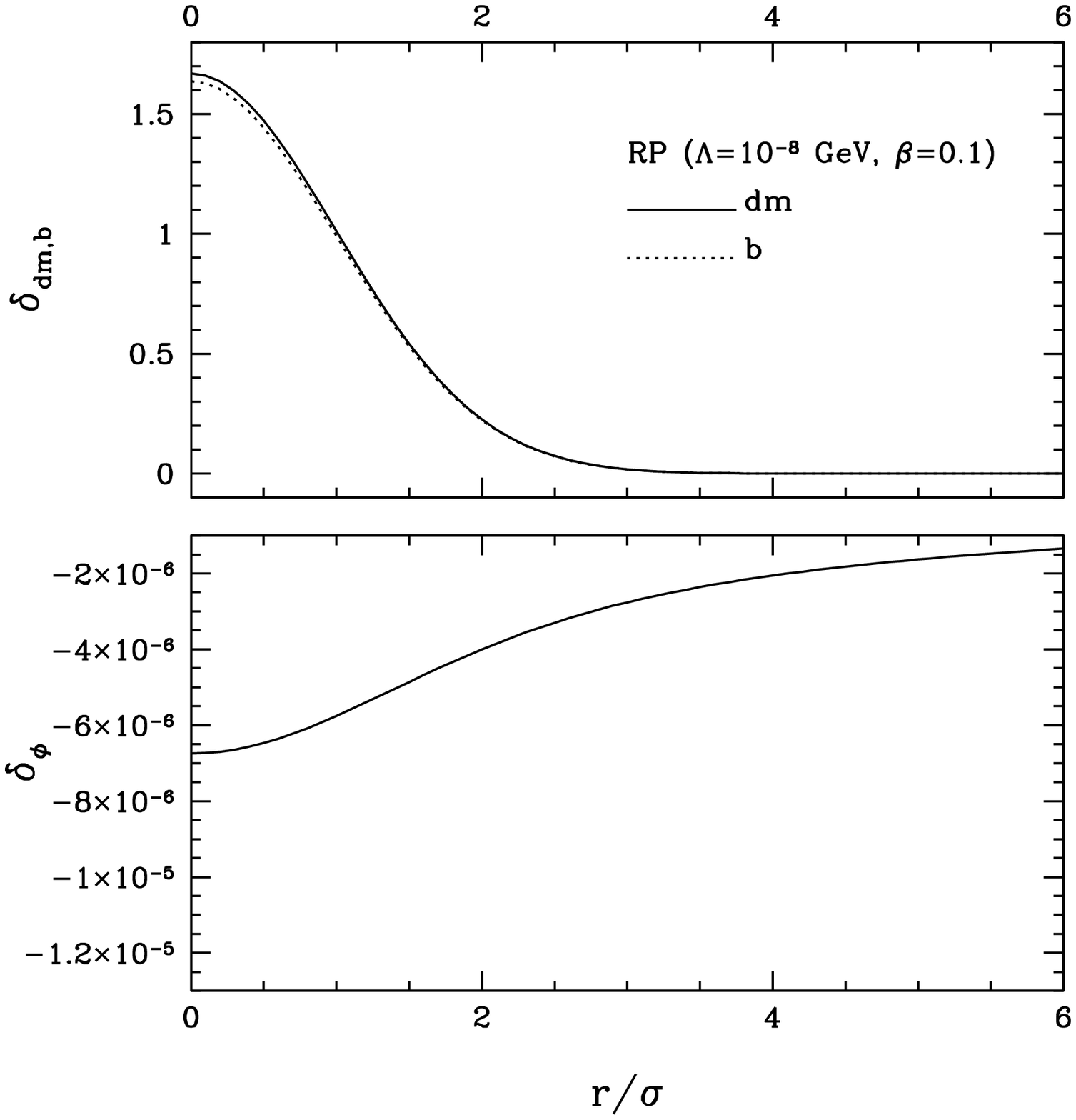}
\caption{Density profiles of DM, baryons (top) and DE (bottom) at the 
present time in the uncoupled (left) and coupled (right) RP model.
Plots are given for the comoving scale $\sigma=20h^{-1}$ Mpc.
The mean matter density contrast within the radius $R=\sigma$ is 
${\bar\delta}=1$ at the present time.} 
\label{fig5}
\end{figure}
\begin{figure}[h!]
\centering
\includegraphics[height=7.cm,angle=0]{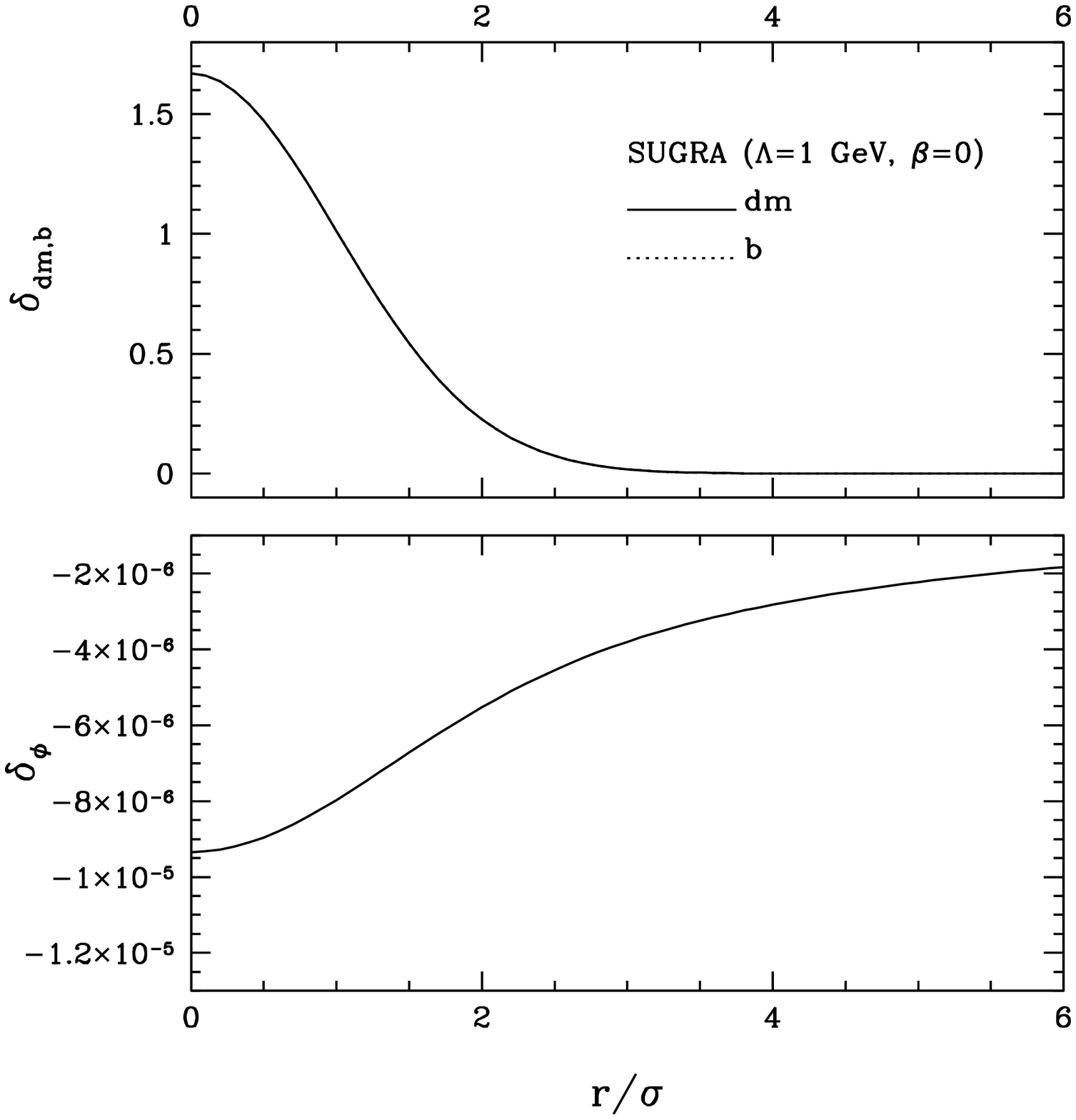}
\includegraphics[height=7.cm,angle=0]{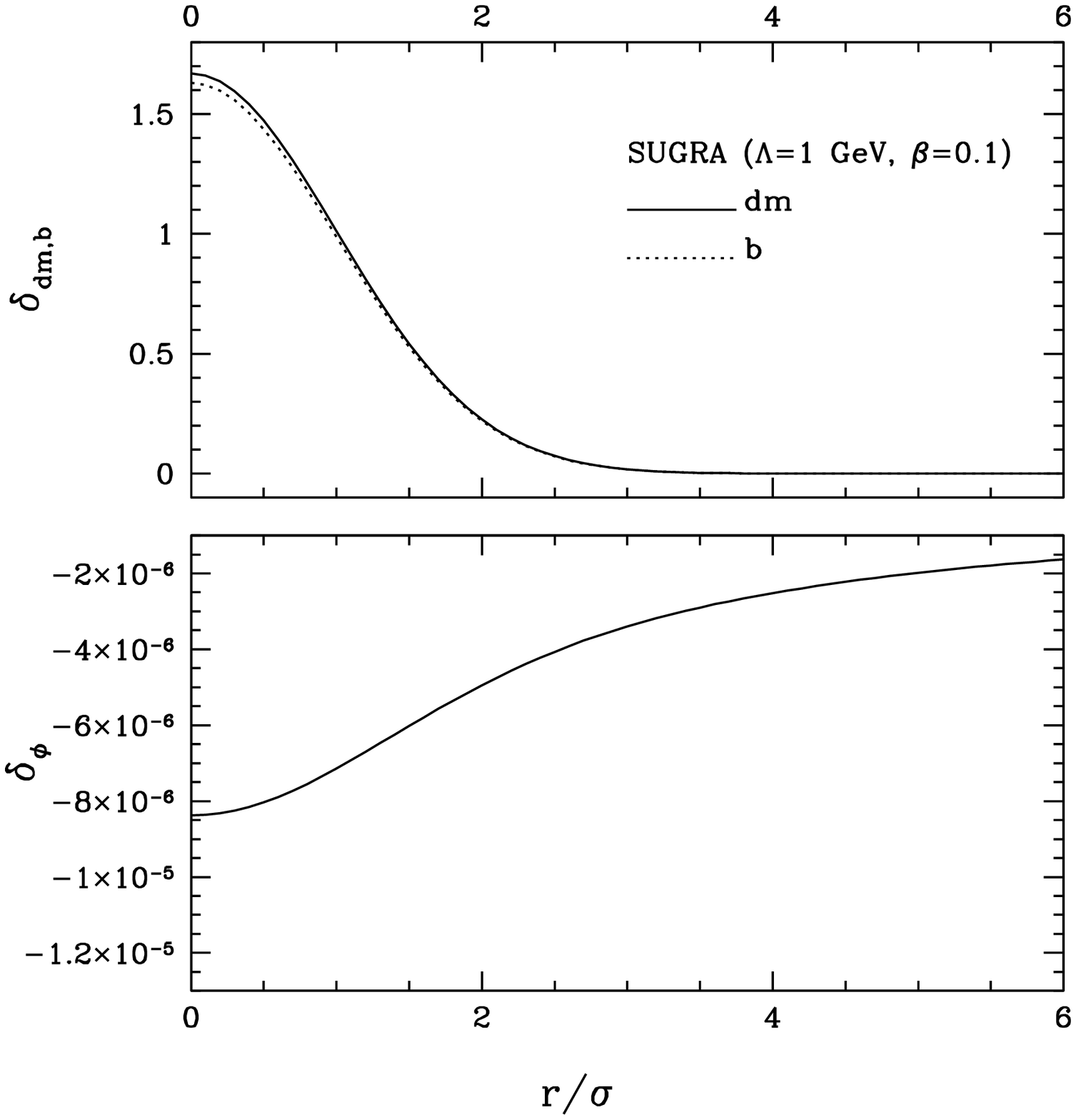}
\caption{Same as Figure \ref{fig5} but using the SUGRA potential.}
\label{fig6}
\end{figure}

Formation of DE overdensities during $\phi$MDE is understood as follows.
Let $\delta w$ be the first order
correction to the state parameter $w$ due to the perturbation in the 
scalar field:
\begin{equation}
\delta w  = {1 \over {\bar\rho}_{\phi}} \left(\delta P_\phi -w \delta\rho_\phi
\right)
\label{delw}
\end{equation}
so that $w+\delta w$ is the perturbed state parameter inside the fluctuation.
Noticing that if ${\bar\rho}_V$ is negligible with respect to ${\bar\rho}_k$,
yielding $w\sim 1$, and $\delta\rho_k>0$, $\delta\rho_V<0$ as previously 
observed, we will also have that 
${\bar\rho}_k + \delta\rho_k >> {\bar\rho}_V + \delta\rho_V$.
Therefore, $w+\delta w \sim w \sim 1 $ (or $\delta w \sim 0$) and from 
(\ref{delw}) it follows $|\delta\rho_k| >> |\delta\rho_V|$.
 
\begin{figure}[h!]
\centering
\includegraphics[height=10.cm,angle=0]{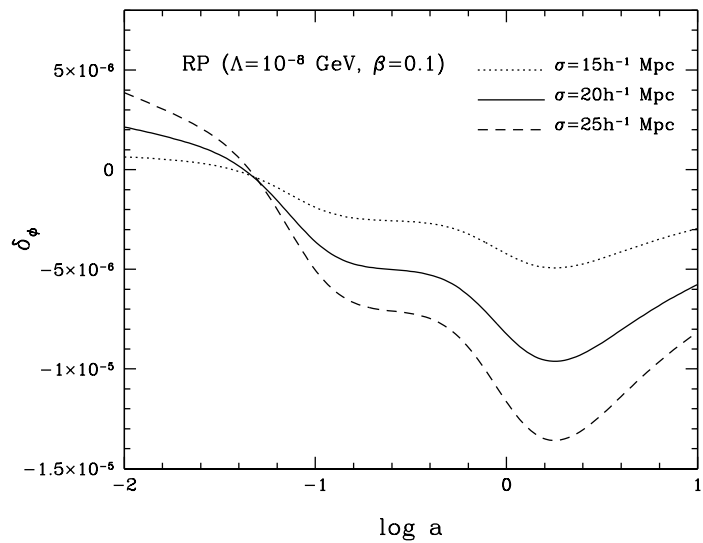}
\vskip -1.1truecm
\caption{Evolution of the DE density contrast, $\delta_\phi$, at the center 
of a spherical matter overdensity in coupled RP models 
when different scales $\sigma$ are considered. Perturbations on 
shorter scales are suppressed.}
\label{fig7}
\end{figure}
\begin{figure}[h!]
\centering
\includegraphics[height=10.cm,angle=0]{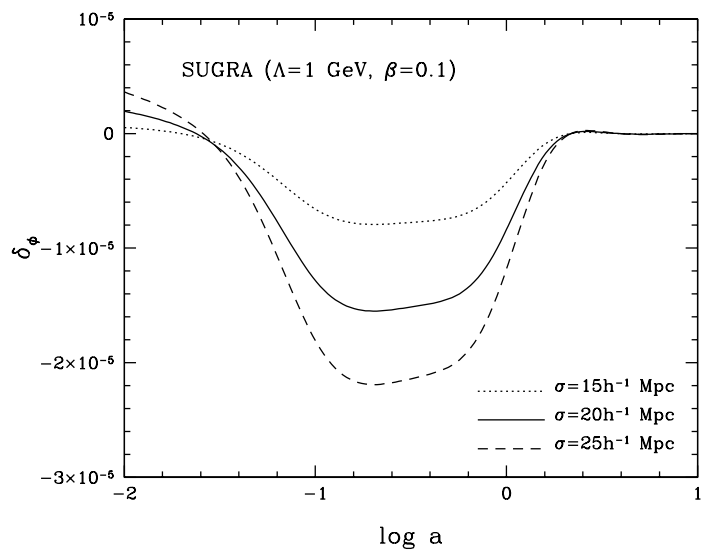}
\caption{Same as Figure \ref{fig7} but using SUGRA potential.}
\label{fig8}
\end{figure}

After $\phi$MDE, $\delta_\phi$  falls off to negative values 
resuming the same qualitative behavior as in the uncoupled case. 
Notice that fluctuation amplitudes decrease when increasing the coupling 
strength. In fact, it shortens the period between the 
end of $\phi$MDE and DE dominance so that less time will be left for the 
perturbation growth.

As soon as DE becomes the dominant component, $|\delta_\phi|$ starts to 
decrease to zero as the Universe approaches the de Sitter attractor.

Figure \ref{fig4} shows the mean density contrast of DM and baryons,
${\bar \delta}_{dm,b}$, within the radius $r=\sigma$, as a function of the 
scale factor. Results are given for the SUGRA model, only slightly differences 
occur when considering different cases. 
As expected, growth of matter perturbations is suppressed once DE dominates the
cosmic expansion. Also, notice how coupling introduces a bias between 
baryons and DM causing fluctuations to grow at different rates. 

Mean density contrasts, ${\bar \delta}_{dm,b}$, are easily obtained from 
({\ref{delmedio}) as the assumed initial gaussian shape of matter perturbations
is kept during their evolution. This is confirmed by numerical results
and shown in the top panels of Figures (\ref{fig5}) and (\ref{fig6}) 
which display the density 
profiles of DM and baryons for different models at the present time.
Density profiles of DE are then shown in the bottom panels of the same figures.

Figures (\ref{fig7}) and (\ref{fig8}) compare the growth of DE perturbations, 
when different scales are considered showing how perturbations 
on shorter scales are suppressed.  

So far, we have considered the behavior of DE perturbations in the presence 
of matter overdensities. We now look at what happens in the presence of 
matter underdensities or voids. It is not surprising to expect that DE perturbations 
behave in an opposite fashion given that local voids increase the local value
of $H$ and consequentely $\delta\phi$, $\dot \delta\phi<0$. This is confirmed 
by the Figures (\ref{fig9}), (\ref{fig10}) and (\ref{fig11}) which display 
the time evolution of $\delta_\phi$ for matter density contrasts and  
scales typical of voids and supervoids (see next section).
Although different models are considered, our results are quite general.
We find the largest DE inhomogeneities 
($|\delta_\phi|\sim {\cal O}(10^{-4}-10^{-3})$) 
to be associated with objects on very large scale ($\sim 300h^{-1}$ Mpc) which 
existence was recently postulated in \cite{inoue} and \cite{rudnick}.
On typical supervoid scales we obtain 
$|\delta_\phi|\sim {\cal O}(10^{-5}-10^{-4})$
while we find even smaller $|\delta_\phi|\sim {\cal O}(10^{-6}-10^{-5})$ in the 
presence of voids and supercluster.

\section{Summary and conclusions}
\label{conclusion}

Unlike cosmological constant models ($\Lambda$CDM), models of DE which have 
a dynamical nature yield a varying state parameter $w(a)$.
Although current limits on $w$ are consistent with a cosmological constant
($w=-1 \pm 0.1$, \cite{riess}), detecting either $w\ne-1$ or its time variation
($dw/da \ne 0$), would provide a crucial support for dynamical DE.
Nevertheless, many models of dynamical DE predict no substantial deviation
from $w=-1$ in the late time evolution providing a background cosmology very 
closed to that of $\Lambda$CDM.

Anyway, dynamical DE stops to mimic a cosmological constant when one deals 
with their clustering properties as dynamical DE is 
expected not to be perfectly homogeneous.
Clustering properties of dynamical DE in the vicinity of matter inhomogeneities
\begin{figure}[h!]
\centering
\includegraphics[height=10.cm,angle=0]{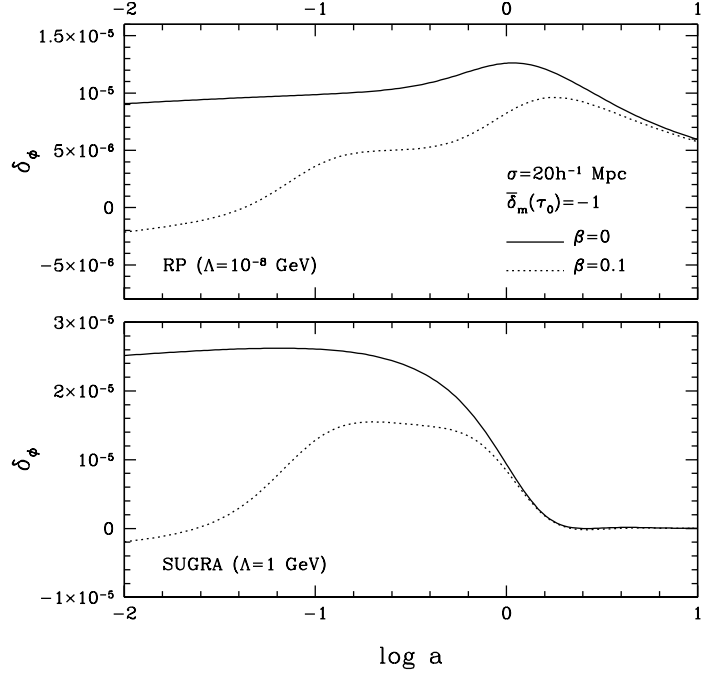}
\vskip -0.8truecm
\caption{Evolution of the DE density contrast, $\delta_\phi$, at the center 
of a spherical matter underdensity in RP (top) and SUGRA (bottom) models.
Plots are given for the comoving scale $\sigma=20h^{-1}$ Mpc.
The mean matter density contrast within the radius $R=\sigma$ is 
${\bar\delta}=-1$ at the present time. Such features are typical of 
matter voids.}
\label{fig9}
\end{figure}
\begin{figure}[h!]
\centering
\includegraphics[height=10.cm,angle=0]{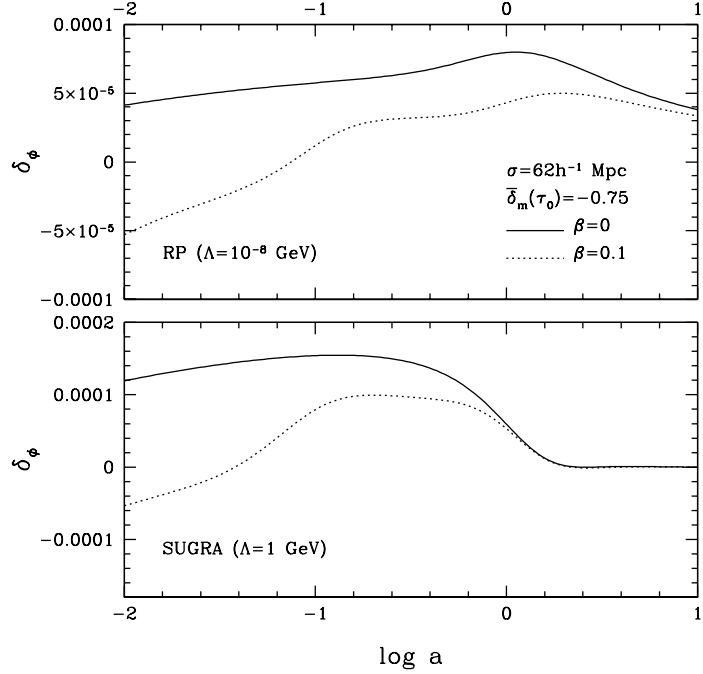}
\vskip -0.8truecm
\caption{Same as Figure \ref{fig9} but for $\sigma=62h^{-1}$ and $
{\bar\delta}=-0.75$. Such values has been estimated for the Bo\"{o}tes 
supervoid (see text).}
\label{fig10}
\end{figure}
\begin{figure}[h!]
\centering
\includegraphics[height=10.cm,angle=0]{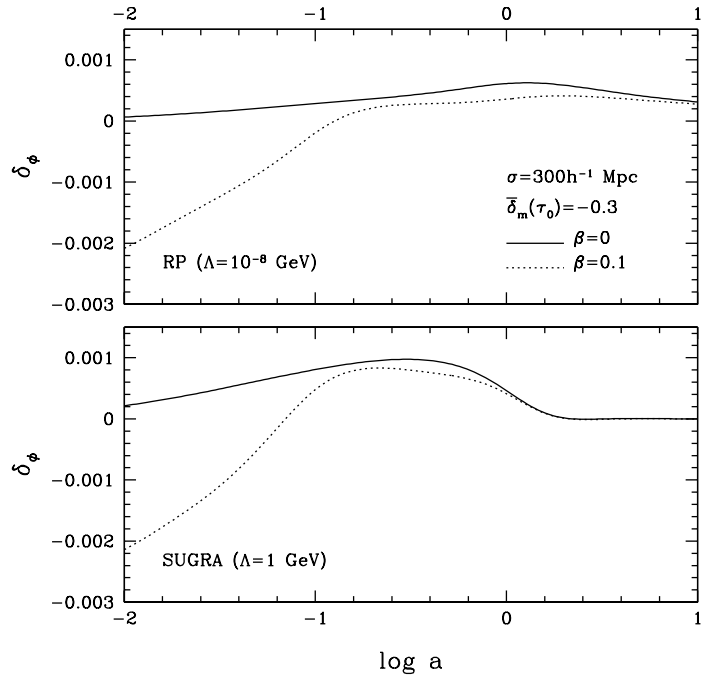}
\caption{Same as Figure \ref{fig9} but for $\sigma=300h^{-1}$ and $
{\bar\delta}=-0.3$. The existence of matter voids with these properties 
has been recently postulated to explain the observed large angle anomalies 
of the CMB (see text).}
\label{fig11}
\end{figure}
was recently studied by Dutta \& Maor \cite{dutta} and  Mota et al \cite{mota}
 using two 
different approaches. While the formers face the problem by numerical methods
evolving the linearized perturbations equations, the latters use an 
analitycal approach which permitted them to extend the analysis even 
to the nonlinear regime.

One of the most striking results of the above works is that DE tends to form 
voids in correspondence of linear matter overdensities. 
However, discrepancies between quantitative results are found. 
When $\delta_m\sim {\cal O} (1)$, Dutta \& Maor \cite{dutta} find significant void amplitudes,
$|\delta_\phi|\sim {\cal O}(10^{-2})$, which could be relevant to observations, 
e.g. on supercluster scales. On the other hand,  Mota et al \cite{mota} report 
$|\delta_\phi|\sim {\cal O}(10^{-5})$ .

In both the works, DE is modeled as a self--interacting scalar field 
minimally coupled to matter and gravity.

In this paper we have extended the analysis to a wider class of DE 
scalar fields admitting tracker solutions also allowing for a possible coupling 
between DM and DE. By using the same numerical approach as Dutta \& Maor \cite{dutta}, we have 
studied the clustering properties of DE in presence of linear
matter inhomogeneities on supercluster, void and supervoid scales.

Superclusters are the largest known gravitationally bound massive
structures with typical radii of about $10-25 h^{-1}$ Mpc and mean density 
contrasts $\bar{\delta}\sim 1-15$. 
Typical examples are the local
supercluster (LSC), of which our galaxy is a part, and the Shapley supercluster
(SSC) which is $\sim 650$ Mlys from us. 
The LSC has a mean overdensity of $\bar{\delta}_m\sim 2-3$ over a 
scale $\sim 15h^{-1}$ Mpc \cite{hoffman} while the SSC has been 
found to have an overdensity
of $\bar{\delta}_m \sim 10.3$ over a scale of $10.1h^{-1}$ Mpc
\cite{bardelli}.
In addition to matter overdensities, Universe contains voids of matter typically
having radii similar to superclusters or even larger (supervoids). 
Data from 2dFGRS are consistent with voids having average radii  
of $\sim 15h^{-1}$ Mpc and average mean density contrast 
$\bar{\delta}_m\sim -0.93$ \cite{hoyle}. 
An example of a supervoid is the Bo\"{o}tes void found to be rougthly spherical
with radius of $\sim 62h^{-1}$ Mpc \cite{kirschner} and a mean density contrast
estimeted to be $-0.8<{\bar\delta}_m<-0.66$.

Recently, the possibility that extremely large 
voids might exist with radii of $\sim 100 -300$ Mpc, has been considered. 
In particular in 
\cite{inoue} it is
shown that the Integrated Sachs Wolfe (ISW) effect due to a void with 
radius $\sim 200-300$ Mpc and 
${\bar\delta}_m \sim-0.3$ would be observed as a cold spot in the 
CMB radiation explaining the 
observed large angle CMB anomalies (see also \cite{rudnick} for a similar 
conclusion).
According to our results (and those of  Mota et al \cite{mota} as well) 
such extremely large objects would  correspond to the largest DE 
overdensities.  

We conclude summarizing our main results concerning the 
behavior of DE perturbations in the vicinity of matter inhomogeneities.
In the presence of matter overdensities we find that:
\begin{enumerate}

\item DE tends to form voids if no coupling to DM is present

\item in coupled models, DE overdensities form during $\phi$MDE.
After this stage $\delta_\phi$ becomes negative resuming
the same behavior as in the uncoupled case. 
\end{enumerate}

If matter underdensities are considered, DE perturbations behave in the 
opposite fashion.
From our analysis we obtain $|\delta_\phi| \sim {\cal O}(10^{-6}-10^{-5})$ 
for voids and supercluster, $|\delta_\phi| \sim {\cal O}(10^{-5}-10^{-4})$ 
for supervoids and if extremely large voids exist 
$|\delta_\phi| \sim {\cal O}(10^{-4}-10^{-3})$. Our results are consistent
with those of Mota et al \cite{mota} indicating that if DE is described
by a scalar field, either uncoupled or coupled with DM, it would be almost 
homogeneous on sub-horizon scales above that of galaxy clusters. 

Further, in general we have:  

\begin{enumerate}

\item
within matter inhomogeneities DE can form voids as well as overdensities.
However, the behavior of $\delta_\phi$ is mainly related to 
how the background DE energy density evolves rather than to 
whether a matter fluctuation is an overdensity or an underdensity

\item DE perturbation growth is sensitive to the scale considered. 
On shorter scales perturbations are suppressed 

\item accelerated expansion yields a suppression of DE inhomogeneities, i.e
$|\delta_\phi|$ decreases to zero as the universe approaches the de Sitter 
attractor. 
\end{enumerate}

We have shown in the previous section that 
$|\delta_\phi|$ decreases at the increasing of the coupling 
strengh $\beta$ when supercluster and void scales are considered. 
According to the first of the above claims, the fluctuation 
suppression due to the coupling is then to impute to the fact that 
perturbations grow on horseback of two distinct evolutionary phases of 
the scalar field, namely $\phi$MDE and usual 
tracking phase in matter era, changing their sign.
It would be interesting to ascertain whether such effect also holds 
when dealing with nonlinear matter collapse, e.g. on galaxy cluster scales.
If so, very small DE perturbations are expected, at most not larger than what
found by Mota et al \cite{mota} when considering nonlinear scales in 
uncoupled DE models, i.e. $\delta_\phi\sim {\cal O}(10^{-5})$.
This would be a quite unexpected result since it seems natural to 
believe that infalling matter, when coupled to DE,
will drag along DE permitting larger DE perturbations.

Suppression of DE perturbations due to the coupling is however 
strictly true only after $\phi$MDE if larger scales are considered.
During $\phi$MDE, DE fluctuations are only marginally suppressed on supervoid 
scales, while they can be as larger as matter fluctuations on extremely large
scales ($\sim200-300$ Mpc). Figure \ref{fig11} indicates $\delta_\phi \sim \delta_m$ at 
$z\sim 100$ and even larger values are expected for higher $z$.
suggesting the possibility that DE clustering might be detected through
the ISW effect.

As already observed in \cite{isw}, an interaction between DM and DE
changes both the scaling of the DM energy density and the growth rate 
of matter perturbations affecting the time evolution of the metric potentials
and, consequentely, the ISW effect. 
Our results point out that, in coupled cosmologies,  
a further contribution to ISW effect can arise during $\phi$MDE
from DE perturbations
associated with very large voids of matter.

Further investigations on this last point and the behavior of DE in 
the presence of nonlinear matter inhomogeneities
are left to future works.

\begin{ack}
I wish to thank A. Gardini, for useful hints and discussions.
I am also grateful to J.R. Kristiansen for reading the manuscript and 
pointing out some corrections.
This work is supported by the Research Council of Norway, project number 162830.
\end{ack}
  
\appendix
\section{}
\label{equations}
In a spacetime described by the metric (\ref{metric}):
\begin{eqnarray}
\fl
	ds^{2}=-dt^2+{\cal U}(t,r)dr^2+{\cal V}(t,r)\left(
	d\theta^2+\sin^{2}\theta d\varphi^2 \right) 
\end{eqnarray}
the Einstein's equations take the form:
\begin{eqnarray}
\fl
\label{metr1}
	\frac{1}{2}\frac{\ddot{\cal U}}{\cal U}+
	\frac{1}{2}\frac{\dot{\cal U}}{\cal U}\frac{\dot{\cal V}}{\cal V}-
	\frac{1}{4}\frac{\dot{\cal U}^2}{{\cal U}^2}+
	\frac{1}{\cal U}\left(\frac{1}{2}\frac{\cal U'}{\cal U}\frac{{\cal V}'}
        {\cal V}+
	\frac{1}{2}\frac{{\cal V}'^2}{{\cal V}^2}-
	\frac{{\cal V}''}{\cal V}\right)=
	-4\pi G \left(T^0_0-T^1_1+2T^2_2 \right) 
\\\fl
	\frac{1}{2}\frac{\ddot{\cal V}}{\cal V}+
	\frac{1}{4}\frac{\dot{\cal U}}{\cal U}\frac{\dot{\cal V}}{\cal V}+
	\frac{1}{\cal V}+
	\frac{1}{2{\cal U}}\left(\frac{1}{2}\frac{{\cal U}'}{\cal U}
        \frac{{\cal V}'}{\cal V}-
	\frac{{\cal V}''}{\cal V}\right) 
	=-4 \pi G\left(T^0_0+T^1_1\right) 
\\\fl
 	\frac{1}{2}\frac{\dot{\cal U}}{\cal U}\frac{\dot{\cal V}}{\cal V}+
        \frac{1}{4}\frac{\dot{\cal V}^2}{{\cal V}^2}+
	\frac{1}{\cal V}+
	\frac{1}{\cal U}\left( \frac{1}{2}\frac{{\cal U}'}{\cal U}
        \frac{{\cal V}'}{\cal V}+
	\frac{1}{4}\frac{{\cal V}'^2}{{\cal V}^2}-
	\frac{{\cal V}''}{\cal V}\right) 
	=-8 \pi G T^0_0
\\\fl
        \frac{1}{2}\frac{\dot{\cal U}}{\cal U}\frac{{\cal V}'}{\cal V}+
	\frac{1}{2}\frac{\dot{\cal V}}{\cal V}\frac{{\cal V}'}{\cal V}-
	\frac{\dot{\cal V}'}{\cal V}
        =-8\pi G T^0_1 
\label{metr4}
\end{eqnarray}
while from (\ref{contin}) it follows:
\begin{eqnarray}
\label{a}
\fl
	\ddot{\phi}+\left(\frac{\dot{\cal V}}{\cal V}+
	\frac{1}{2}\frac{\dot{\cal U}}{\cal U}\right)\dot{\phi}-
        \frac{1}{\cal U}\left[\left(\frac{{\cal V}'}{\cal V}-
        \frac{1}{2}\frac{{\cal U}'}{\cal U}\right)\phi'+
        \phi'' \right]+
	\frac{dV}{d\phi}
	=C\rho_{dm}
\\\fl
	\dot{\rho}_{dm}+
	\left[\frac{\dot{\cal V}}{\cal V}+
	\frac{1}{2}\frac{\dot{\cal U}}{\cal U}+
        \left(\frac{{\cal V}'}{\cal V}+
        \frac{1}{2}\frac{{\cal U}'}{\cal U}\right)v^r_{dm}+{v^r_{dm}}'\right]
        \rho_{dm} +\rho_{dm}' v^r_{dm} =-C\dot\phi \rho_{dm}
\\\fl
        \dot  v^r_{dm}+ \left(\frac{\dot{\cal V}}{\cal V}+
	\frac{3}{2}\frac{\dot{\cal U}}{\cal U} + 
        \frac{\dot{\rho}_{dm}}{\rho_{dm}} \right)v^r_{dm}= C\frac{\phi'}{\cal U}
\\\fl
	\dot{\rho}_b+
	\left[\frac{\dot{\cal V}}{\cal V}+
	\frac{1}{2}\frac{\dot{\cal U}}{\cal U}+
        \left(\frac{{\cal V}'}{\cal V}+
        \frac{1}{2}\frac{{\cal U}'}{\cal U}\right)v^r_b+{v^r_b}'\right]
        \rho_b +\rho_b' v^r_b =0
\\\fl
        \dot  v^r_b+ \left(\frac{\dot{\cal V}}{\cal V}+
	\frac{3}{2}\frac{\dot{\cal U}}{\cal U} + 
        \frac{\dot{\rho}_b}{\rho_b} \right)v^r_b= 0
\label{b}
\end{eqnarray}
Here, overdots and primes denote the derivatives with respect to $t$
and the radial coordinate $r$ respectively. 
Only radial motions are considered and terms quadratic 
in $v^r_{dm,b}$ has been neglected as DM and baryons are 
non--relativistic components.

According to (\ref{redef}) and (\ref{split}) we decompose our variables
in an homogeneous part and a perturbation.
It is then straightforward to obtain the equations for the background:
\begin{eqnarray}
\fl\label{eqback1}
  3{\cal H}^2= 8\pi G \left[{\bar\rho}_{dm}+ {\bar\rho}_b+ 
{\bar\rho}_\phi\right]\\\fl
\label{eqback2}
  \ddot{\bar\phi}+3 {\cal H}\dot{\bar\phi}+{\bar V'}= C{\bar\rho}_{dm} \\
\fl\label{eqback3}
 \dot{\bar\rho}_{dm}+3{\cal H}{\bar\rho}_{dm}= 
-C{\bar\rho}_{dm}\dot{\bar \phi} \\\fl
  \dot{\bar\rho}_{b}+3{\cal H}{\bar\rho}_{b}=0 
\label{eqback4}
\end{eqnarray}
To linear order, (\ref{metr1})--(\ref{metr4}) gives 
the equations for the metric perturbations:
\begin{eqnarray}
\fl \label{1}
	\ddot{\zeta}+2H\left(2\dot{\zeta}+\dot{\psi}\right)+
	\frac{2}{a^2}\left( \frac{\zeta'}{r}-
	\frac{2\psi'}{r}-\psi''\right)
	=4 \pi G \left(\delta\rho_{dm}+\delta\rho_b+
        \delta\rho_\phi- \delta P_\phi
	\right)
\\\fl   \label{2}
 	\ddot{\psi}+H\left(5\dot{\psi}+\dot{\zeta}\right)+
	\frac{1}{a^2}\left(\frac{2\zeta}{r^2}-
	\frac{2\psi}{r^2}+\frac{\zeta'}{r}-
	\frac{4\psi'}{r}-\psi''\right) 
	=4\pi G\left(\delta\rho_{dm}+\delta\rho_b+
        \delta\rho_\phi- \delta P_\phi
	\right) \nonumber\\
\\\fl    \label{3}
	2H\left(\dot{\zeta}+2\dot{\psi}\right)+
	\frac{2}{a^2}\left(\frac{\zeta}{r^2}-
	\frac{\psi}{r^2}+\frac{\zeta'}{r}-
	\frac{3\psi'}{r}-\psi''\right)
	=8\pi G \left(\delta\rho_{dm}+\delta\rho_b+ \delta\rho_\phi
	\right)
\\\fl
	\frac{2}{r}\left(
	\dot{\zeta}-\dot{\psi}\right)-2\dot{\psi}'
	=-8\pi G\left[{\bar \rho}_{dm} v^r_{dm} + 
        {\bar \rho}_b v^r_b + 
        \left({\bar \rho}_\phi+{\bar P_\phi}\right) v^r_\phi \right]a^2
\end{eqnarray}
while perturbation equations for DE, DM and baryons follow from 
(\ref{a})--(\ref{b}):
\begin{eqnarray}
\fl
	\delta\ddot{\phi}+3H\delta\dot{\phi}-
	\frac{1}{a^2}\left(\delta\phi''+
	\frac{2}{r}\delta\phi'\right)+ \frac{d^2{\bar V}}{d\phi^2} \delta\phi+
        \left(\dot{\zeta}+2\dot{\psi}\right)\dot{\phi}=C\delta\rho_{dm}
\label{p1}
\\\fl
        \delta\dot{\rho}_{dm}+3H\delta{\rho}_{dm}+
	\left(\dot{\zeta}+2\dot{\psi}\right)\rho_{dm}+ 
        \left({v^r_{dm}}'+\frac{2}{r}v^r_{dm}\right){\rho}_{dm}=
        -C\left(\dot{\bar\phi}\delta\rho_{dm} + 
        \dot{\delta\phi}{\bar\rho}_{dm}\right)
\\\fl
        \dot v^r_{dm} + 2H v^r_{dm}= C\left(\dot \phi v^r_{dm}+ 
        \frac{\delta\phi'}{a^2}\right)
\\\fl
        \delta\dot{\rho}_b+3H\delta{\rho}_b+
	\left(\dot{\zeta}+2\dot{\psi}\right)\rho_b+ 
        \left({v^r_b}'+\frac{2}{r}v^r_b\right){\rho}_b=0
\\\fl
        \dot v^r_b + 2H v^r_b= 0
\end{eqnarray}
Combining equations (\ref{1}), (\ref{2}) and (\ref{3}) gives:
\begin{eqnarray}
\fl
    \left(
	\ddot{\zeta}+2\ddot{\psi} \right)
    +2H\left(\dot{\zeta}+
	2\dot{\psi}\right)
	= -4\pi G\left(\delta\rho_{dm}+\delta\rho_b+ \delta\rho_\phi +
        3\delta P_\phi \right)
\label{combi}
\end{eqnarray}
It is clear, from the above equations, 
that $\chi = \dot{\zeta} +2\dot{\psi}$ is the only combination
which is relevant for the evolution equations of DE, DM and baryons
perturbations.

When rewritten in terms of the
conformal time $\tau$ and the variables $\chi$,
$\delta_i=\delta\rho_i /{\bar\rho}_i$ and
$\theta_i = {v^r_i}' +2v^r_i/r$ ($i=dm,b,\phi$) 
equations (\ref{eqback1})--(\ref{eqback4}) and
(\ref{p1})--(\ref{combi}) gives (\ref{eqpert}) and (\ref{eqback})
respectively.

{}
\end{document}